\documentclass[aps,pre,10pt,showpacs,twocolumn,nofootinbib,floatfix]{revtex4-1}

\usepackage{graphicx,amsmath,amssymb}
\usepackage[usenames]{color}
\usepackage[latin1]{inputenc}
\usepackage{ulem}

\usepackage{multirow}

\newcommand{\eq}[1]{Eq.~(\ref{#1})}
\newcommand{\fig}[1]{Fig.~\ref{#1}}
\newcommand{\tab}[1]{Table~\ref{#1}}

\newcommand{\avg}[1]{\langle #1 \rangle}
\newcommand{\olcite}[1]{Ref.~\onlinecite{#1}}
\newcommand{\ahum}[1]{``#1''}

\newcommand{\PA}{P(A|T,\Pi)}
\newcommand{\Tc}{T_{\rm c}}

\begin{document}

%% 87.16.D-	Membranes, bilayers, and vesicles
%% 87.14.Cc 	Lipids
%% 64.70.-p	Specific phase transitions
%% 82.20.Wt 	Computational modeling; simulation 


\title{The main transition in the Pink membrane model: finite-size scaling and 
\\ the influence of surface roughness}

\author{Sina Sadeghi and R. L. C. Vink}
\affiliation{Institute of Theoretical Physics, Georg-August-Universit\"at 
G\"ottingen, Friedrich-Hund-Platz~1, D-37077 G\"ottingen, Germany}

\begin{abstract} We consider the main transition in single-component membranes 
using computer simulations of the Pink model [D.~Pink {\it et al.},~Biochemistry 
{\bf 19}, 349 (1980)]. We first show that the accepted parameters of the Pink 
model yield a main transition temperature that is systematically below 
experimental values. This resolves an issue that was first pointed out by 
Corvera and co-workers [Phys.~Rev.~E {\bf 47}, 696 (1993)]. In order to yield 
the correct transition temperature, the strength of the van der Waals coupling 
in the Pink model must be increased; by using finite-size scaling, a set of 
optimal values is proposed. We also provide finite-size scaling evidence that 
the Pink model belongs to the universality class of the two-dimensional Ising 
model. This finding holds irrespective of the number of conformational states. 
Finally, we address the main transition in the presence of quenched disorder, 
which may arise in situations where the membrane is deposited on a rough 
support. In this case, we observe a stable multi-domain structure of gel and 
fluid domains, and the absence of a sharp transition in the thermodynamic limit. 
\end{abstract}

\maketitle

\section{Introduction}

Lipid membrane bilayers are abundant in nature and to understand their 
properties is of paramount importance \cite{citeulike:1022291, 
citeulike:3042594, citeulike:416456}. One aspect that has received much 
attention are collective phenomena (phase transitions) taking place in these 
systems. Among the different phase transitions that can 
occur~\cite{citeulike:10527301, citeulike:6218761, citeulike:6593910, 
citeulike:3850753}, the main phase transition is presumably the most important 
and well studied one~\cite{citeulike:7028415, citeulike:10527551}. This 
transition, typically driven by the temperature $T$, is between a \ahum{gel} and 
a \ahum{fluid} phase. At low $T$, the bilayer is in the gel phase (characterized 
by nematic chain order of the lipid tails), while at high $T$ the bilayer 
assumes the fluid phase (characterized by the absence of nematic chain order).

Computer simulations have become a well established tool to model the main 
transition. The challenge in simulations is to strike a balance between the 
level of detail to include, and the time and length scale one wishes to address 
\cite{citeulike:1007006}. Since collective phenomena involve many molecules and 
entail large length scales it is clear that, in order to describe the main 
transition, a significantly coarse-grained particle model is crucial. Strictly 
speaking, one needs to address the thermodynamic limit (infinite particle 
number) since only there phase transition properties become properly defined. 
Indeed, the need for coarse grained modeling of lipid bilayers is well 
recognized~\cite{citeulike:8529576, citeulike:7152547, citeulike:270315}.

An early and highly successful coarse grained approach to study the main 
transition has been the particle model introduced by David Pink and 
co-workers~\cite{Pink1980, springerlink:10.1007/BF01295077, 
doi:10.1139/p80-083}. In this model, the so-called Pink model, only the {\it 
orientational} degrees of freedom of the hydrophobic lipid tails are included, 
while the {\it positional} degrees of freedom of the hydrophilic heads are 
disregarded. This model, due to its simplicity, allows for the investigation of 
very large systems, and the nature of the main transition can be probed in great 
detail. Indeed, key features of the main transition in the Pink model compare 
well to experiments~\cite{citeulike:9172835}.

However, despite the great success the Pink model has enjoyed, there remain some 
open questions. In \olcite{Laradji}, it was noted that the Pink model at the 
experimentally determined transition temperature does {\it not} undergo any 
transition. While in systems of finite size there were indications of a 
transition, these vanished in larger systems. This raises the question as to why 
no transition could be detected. The aim of this paper is to resolve this issue. 
As it turns out, to properly model the main transition, a finite-size scaling 
study is essential. Computer simulations inevitably deal with only a finite 
number of particles, and their output will depend on the number of particles 
used, especially near phase transitions. Finite-size scaling provides the 
framework to systematically extrapolate simulation data to the thermodynamic 
limit. To date, finite-size scaling studies of the Pink model are scarce, with 
\olcite{Laradji} being a notable exception. The present paper aims to fill this 
gap. Our main finding is that, in order to observe the main transition in the 
Pink model at experimentally relevant temperatures, one of the model parameters 
needs to be adjusted. This follows quite naturally when one realizes that the 
universality class of the Pink model is just the one of the two-dimensional (2D) 
Ising model~\cite{springerlink:10.1007/BF01295077}. As we will show for three 
lipid species, the \ahum{standard} Pink model parameters yield a critical 
temperature distinctly below the experimental main transition temperature. 
Consequently, a \ahum{re-tuning} of the standard Pink parameters is urgently 
needed.

As an application, we also address the fate of the main transition in the 
presence of quenched (immobilized) impurities using the Pink model. The 
experimental motivation to do so is that this situation may resemble that of a 
membrane supported on a rough substrate. In binary lipid mixtures, the effect of 
such impurities on lateral phase separation has recently attracted much 
attention \cite{citeulike:6599228, citeulike:7115548, citeulike:8864903, 
citeulike:9134200, citeulike:8610600, citeulike:10648601}. In this paper, we 
present simulation results for the corresponding scenario in a single component 
bilayer undergoing the main transition. Within the framework of the Pink model, 
we find that quenched impurities prevent the main transition from taking place, 
already at low impurity concentrations. Instead of the formation of macroscopic 
gel and fluid domains, we now obtain a stable multi-domain structure, which 
strikingly resembles experimental results. The theoretical justification is that 
the impurities induce a change in universality toward the 2D {\it random-field} 
Ising class. As is well known, the latter does not support an order-disorder 
phase transition in the thermodynamic limit~\cite{imry.ma:1975, imbrie:1984, 
bricmont.kupiainen:1987, aizenman.wehr:1989}.

\section{The Pink model}

In the Pink model, the lipid bilayer is assumed to consist of two independent 
monolayers. Each monolayer is represented by a triangular 2D lattice consisting 
of $N$ sites, and each lattice site contains a single lipid chain. Each lipid 
molecule is comprised of two independent hydrophobic acyl chains and a 
hydrophilic polar head. The polar heads are translationally frozen to the 
lattice, and no particular structure for the polar head groups is assumed. The 
only degrees of freedom included in the Pink model are the acyl chain 
conformations. These are not simulated directly (i.e.~one does not explicitly 
model the carbon atoms) but are captured in a coarse-grained fashion whereby the 
chain conformations are grouped into $\alpha=1,\ldots,q$ discrete states. The 
original Pink model uses $q=10$, but we will consider different values also. 
These states include the ground state $(\alpha=1)$, eight low-energy excitations 
$(\alpha=2,\ldots,q-1)$, while all remaining conformations are grouped into a 
single disordered state $(\alpha=q)$. Each state $\alpha$ is characterized by 
three coarse-graining parameters, namely an internal energy~$E_\alpha$, a 
cross-sectional area~$A_\alpha$, and a degeneracy~$D_\alpha$ counting the number 
of chain conformations with energy $E_\alpha$ and area $A_\alpha$.

\begin{figure}
\begin{center}
\includegraphics[width=\columnwidth]{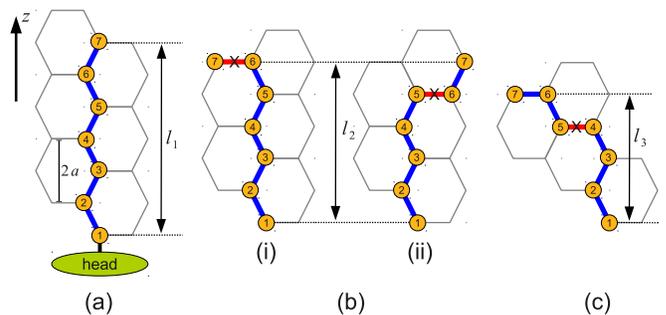} 
\caption{\label{fig1} Typical chain conformations of the Pink model with $M=7$, 
showing (numbered) carbon atoms placed on the nodes of a hexagonal lattice. The 
atom connected to the head group is labeled $i=1$ but for clarity the head group 
is only drawn for the ground state. The $z$-direction indicates the bilayer 
normal, while the vertical double-arrows indicate the projected length. (a) The 
ground state $\alpha=1$, consisting of only {\it trans} bonds. (b) The two 
conformations that constitute the first excited state $\alpha=2$ containing one {\it 
gauche} bond (marked with a cross). (c) Conformation belonging to the second 
excited state $\alpha=3$. The internal energy is the same as in (b) but the projected 
length is shorter (the other $\alpha=3$ conformation has the {\it gauche} bond 
between atoms $3-4$).}
\end{center}
\end{figure}

\subsection{coarse graining parameters}

To determine the coarse graining parameters, we assume that a single acyl chain 
consists of $i=1,\ldots,M$ carbon atoms, thereby containing $M-1$ carbon-carbon 
bonds, and that bonds are either in a {\it trans} or {\it gauche} configuration. 
The {\it trans} configuration yields the lowest energy, while the {\it gauche} 
configuration has a slightly higher energy. The energy difference between the 
{\it trans} and {\it gauche} configuration is denoted $\Gamma$ 
(\tab{state_summary}). To understand the difference in geometry between {\it 
trans} and {\it gauche} bonds consider a chain segment of four consecutive 
carbon atoms. The positions of the first three atoms define a two-dimensional 
plane. In the {\it trans} configuration, the fourth atom remains in the plane, 
while in the {\it gauche} configuration, it leaves the plane, and it can do so 
inward or outward. Thus, each {\it gauche} bond is two-fold degenerate. In the 
Pink model, it is assumed that each $2n$-th {\it gauche} bond takes the chain 
back to the original plane, and so the {\it gauche} degeneracy is given by
\begin{equation}\label{eq:dg}
 G = 2^{{\rm ceil}(n/2)} \quad,
\end{equation}
where $n$ denotes the total number of {\it gauche} bonds in the chain, and where 
the function $\rm ceil$ means \ahum{rounding-up} to the nearest integer.

It is convenient to {\it mathematically} represent the chain conformations on a 
hexagonal lattice with next-nearest neighbor distance $2a$. We emphasize that 
this lattice is merely an aid to identify the low energy chain conformations 
which are needed to set the coarse-graining parameters: it should not be 
confused with the triangular simulation lattice on which the Pink Hamiltonian 
will eventually be defined. The carbon atoms are placed on the nodes of the 
hexagonal lattice following certain rules, and nearest-neighbor connections 
between atoms represent carbon-carbon bonds. The ground state $\alpha=1$ 
corresponds to the chain conformation that is maximally stretched 
[\fig{fig1}(a)]. Note that, in the ground state, the atoms are alternatingly 
placed on the left and right lattice node, yielding a characteristic 
\ahum{zig-zag} pattern. The ground state by definition contains only {\it trans} 
bonds, its internal energy is set to zero as a reference $E_1=0$, and it is 
obviously non-degenerate $D_1=1$. The cross-sectional area of the ground state 
has experimentally been determined as $A_1=20.4$~\AA$^2$ \cite{Pink1980}. We 
also introduce the projected length $l$ of the conformation, defined as the 
difference in $z$-coordinate between the carbon atom closest to the head group 
$(i=1)$ and the one furthest away $(i=M)$, with the $z$-direction as indicated 
in the figure. For the ground state, it follows that $l_1 = (M-1)a$.

The eight low-energy excitations ($\alpha=2,\ldots,9$) are obtained by 
systematically incorporating {\it gauche} bonds. The effect of such a bond is to 
disrupt the \ahum{zig-zag} pattern of the ground state. That is, one no longer 
places the atoms alternatingly on left and right nodes, but also allows for 
\ahum{excursions} whereby for two consecutive atoms the same direction is 
chosen. Each such excursion corresponds to a {\it gauche} bond, and has energy 
cost $\Gamma$. The {\it gauche} bonds are introduced according to the following 
rules: (1) The two bonds in the chain closest to the head group must always be 
in the {\it trans} configuration. In \fig{fig1}, these correspond to the bonds 
between atoms $1-2$ and $2-3$. (2) At most three {\it gauche} bonds are allowed, 
and each time such a bond is included there is an energy cost $\Gamma$. (3) The 
projected chain length $l$ must obey $l_1-l \leq 3a$. (4) The acyl chain cannot 
fold back onto itself. In the coordinate system of \fig{fig1}, this means that 
the $z$-coordinates of the atoms must obey $z_{i+1} \geq z_i$.

Following these rules, we show in \fig{fig1}(b) the chain conformations (i) and 
(ii) that form the first excited state $\alpha=2$. In (i), a single {\it gauche} 
bond is placed at the very chain end, while in (ii) it is placed at the 
second-last position. One immediately sees that both conformations have the same 
energy $E_2=\Gamma$, and the same projected length $l_2=(M-2)a$. To compute the 
cross-sectional area, one assumes volume conservation for the lipid chains: 
$A_\alpha l_\alpha = A_1 l_1$. Hence, from the (known) ground state values, the 
cross-sectional area of the excited state follows. Note that, by placing the 
{\it gauche} bond at the third-last position [\fig{fig1}(c)], a shorter 
projected length is obtained, and so conformation (c) does {\it not} belong to 
the first excited state (even though it has the same energy). The total 
degeneracy of the first exited state $D_2=4$, which is the total number of 
conformations, multiplied by the {\it gauche} degeneracy of \eq{eq:dg}. The 
coarse-graining parameters of the remaining excited states can be found 
analogously, and are listed for completeness in \tab{state_summary}. Finally, 
for the completely disordered state $\alpha=q=10$, one assumes $E_{10} = (0.42 M 
-3.94) \times 10^{-13}$~erg, $A_{10}=34$~\AA$^2$, and degeneracy 
$D_{10}=6\times3^{M-6}$, which have their origins in experimental 
considerations~\cite{doi:10.1139/p80-083}.

\begin{table}
\footnotesize
\begin{ruledtabular}
\begin{tabular}{c c c c c c c c c c c c c c c c}
\multicolumn{4}{c}{state~($\alpha$)} &&& $E_\alpha$  &&& $l_\alpha$ &&& 
$D_\alpha$ &&& \\ \hline \hline
ground state &&& $1$     &&& $0$    &&& $M-1$ &&& $1$ &&& \\ 
\multirow{8}{*}{kink $\left \{ \begin{matrix} \\ \\ \\ \\ \\ \\ \\ \\ \end{matrix} \right.$}   
      &&& $2$  &&& $\Gamma$  &&& $M-2$ &&& $4$ &&& \\
      &&& $3$  &&& $\Gamma$  &&& $M-3$ &&& $4$ &&& \\
      &&& $4$  &&& $\Gamma$  &&& $M-4$ &&& $4$ &&& \\
      &&& $5$  &&& $2\Gamma$ &&& $M-2$ &&& $2(M-6)$ &&& \\
      &&& $6$  &&& $2\Gamma$ &&& $M-3$ &&& $2(M-8)$ &&& \\
      &&& $7$  &&& $2\Gamma$ &&& $M-4$ &&& $2(M-10)$ &&& \\
      &&& $8$  &&& $3\Gamma$ &&& $M-3$ &&& $8(M-8)$ &&& \\
      &&& $9$  &&& $3\Gamma$ &&& $M-4$ &&& $16(M-10)$ &&& \\
disordered &&& $10$ &&& $E_{10}$  &&& $l_1A_1/A_{10}$ &&& $6\times3^{M-6}$ &&& \\
\end{tabular}
\end{ruledtabular}
\caption{\label{state_summary} The coarse graining parameters used to describe 
the acyl chain conformations in the $q=10$ Pink model~\cite{Pink1980, 
springerlink:10.1007/BF01295077, doi:10.1139/p80-083, mouritsen:com-sim}. For 
each state conformation $\alpha$, we list the internal energy $E_\alpha$, the 
projected length $l_\alpha$, and the degeneracy $D_\alpha$. The energy of a 
single {\it gauche} bond equals $\Gamma = 0.45 \times 10^{-13} \, \rm erg$, 
while $M$ denotes the number of carbon atoms in the chain.}
\end{table}

\subsection{Pink model Hamiltonian}

Having specified the coarse-graining parameters, the Hamiltonian of the Pink 
model can be written as~\cite{Mouritsenbook}
\begin{equation}\label{1}
 \mathcal{H}_{\rm Pink} = \mathcal{H}_0 + 
 \mathcal{H}_{\rm VDW} + \mathcal{H}_{\rm P} \quad.
\end{equation}
The first term is the total internal energy of the acyl chains $\mathcal{H}_0 = 
\sum_{i=1}^N E_{s(i)}$, with the sum over all $N$ sites of the triangular 
lattice, and $s(i) \in \{1,\ldots,q \}$ the conformational state at the $i$-th 
lattice site. The second term represents the anisotropic van der Waals 
interaction between adjacent acyl chains $\mathcal{H}_{\rm VDW} = -J_0 
\sum_{\avg{i,j}} I_{s(i)} \, I_{s(j)}$, with $J_0$ the van der Waals coupling 
constant, and $\avg{i,j}$ a sum over all $3N$ nearest-neighboring sites on the 
triangular lattice. The precise value of $J_0$ depends on the chain length, and 
explicit expressions are provided elsewhere~\cite{Pink1980, 
springerlink:10.1007/BF01295077, Ipsen1990}. However, it has been noted that 
these parameters do not always yield a main transition at the expected 
temperature~\cite{Laradji}, and so we will also propose our own values later on. 
The (dimensionless) variables $I_{\alpha}$ measure nematic chain order, and can 
be expressed in terms of the cross-sectional areas \cite{doi:10.1139/p80-083, 
Laradji}
\begin{equation}
 I_{\alpha} = {\omega}_{\alpha} \left ( \frac{9}{5}\frac{A_1}{A_\alpha}- 
 \frac{4}{5} \right ) \left ( {\frac{A_1}{A_\alpha}} \right )^{5/4} \quad, 
\end{equation} 
where $\omega_{10}=0.4$ for the disordered state $\alpha=10$, and 
$\omega_\alpha=1$ otherwise. 

The last term in the Hamiltonian accounts for the interaction between the 
hydrophilic polar head groups and between them and water and also steric 
interactions from both head groups and the lipid chains. Although it is possible 
to consider a more realistic pairwise interaction between the 
headgroups~\cite{Mouritsenbook}, this interaction can be approximated with a 
simple pressure term $\mathcal{H}_{\rm P} = \Pi A$, where $\Pi$ is an effective 
lateral pressure acting on the lipid chains in the bilayer membrane, and $A$ the 
total cross-sectional area occupied by the lipids chains
\begin{equation}\label{eq:a}
 A = \sum_{i=1}^N A_{s(i)} \quad.
\end{equation}

\section{Monte Carlo methods}

To study the phase behavior of the Pink model, we use the Monte Carlo (MC) 
simulation method. We mostly use triangular lattices of size $N=L \times L$ with 
periodic boundary conditions. The principal MC move consists of randomly picking 
one of the lattice sites, read-out the conformational state $\alpha$ of that 
site, and propose a new state $\beta$ drawn randomly from the set of $q$ 
possible states. The new configuration is accepted with the Metropolis criterion
\begin{equation}
 P_{\rm acc}(\alpha \to \beta) = \min \left[1, \frac{D_\beta}{D_\alpha} \exp 
 \left( - \frac{\Delta \cal H}{k_{\rm B} T} \right) \right] \quad,
\end{equation}
where $D$ denotes the state degeneracy, $\Delta \cal H$ the energy difference 
between initial and final configuration as given by \eq{1}, $k_{\rm B}$ the Boltzmann 
constant, and $T$ the temperature. The degeneracy compensates for the fact that 
some of the states have a much larger entropy, and should therefore appear more 
often in the ensemble average.

By virtue of the MC move, the total projected area $A$ given by \eq{eq:a} 
fluctuates during the course of the simulation. In fact, $A$ plays the role of 
order parameter since it changes abruptly at the main phase transition. 
Hence, it is instructive to measure the distribution $\PA$, defined as the 
probability to observe a configuration with projected area $A$. The distribution 
depends on the imposed temperature and pressure, as well as on the linear 
extension $L$ of the triangular simulation lattice. At the main transition, 
$\PA$ assumes a characteristic bimodal shape, from which a number of important 
phase properties are obtained (explicit examples are provided in the next 
section). We note that even with very long simulation runs, distributions $\PA$ 
of high statistical quality are difficult to obtain, especially in the vicinity 
of the main transition. The reason is related to free energy barriers that 
arise from the formation of interfaces~\cite{citeulike:7134501, 
citeulike:4503186, binder:1982}. To overcome this problem, we combine our MC 
simulations with a biased sampling scheme called successive umbrella sampling 
\cite{MarcusMullerSUS}; the latter ensures that $\PA$ is sampled accurately over 
the entire (specified) range in $A$ of interest. A final ingredient to economize 
simulation time is the use of histogram reweighting \cite{HistoRew}. A single 
simulation run yields $\PA$ at a given temperature $T$ and effective pressure 
$\Pi$; histogram reweighting enables us to extrapolate the measured distribution 
to different values $T',\Pi'$. For example, extrapolations in the pressure are 
performed using $P(A|T,\Pi') \propto \PA \, e^{ -(\Pi'-\Pi)A / k_{\rm B}T }$. 
Extrapolations in the temperature can be performed analogously, but also require 
storage of the energy histograms; for implementation details see 
\olcite{citeulike:7554917}.

\section{Results}

\subsection{the \ahum{standard} Pink model revisited}

\begin{figure}
\begin{center}
\includegraphics[width=0.4\textwidth,angle=0]{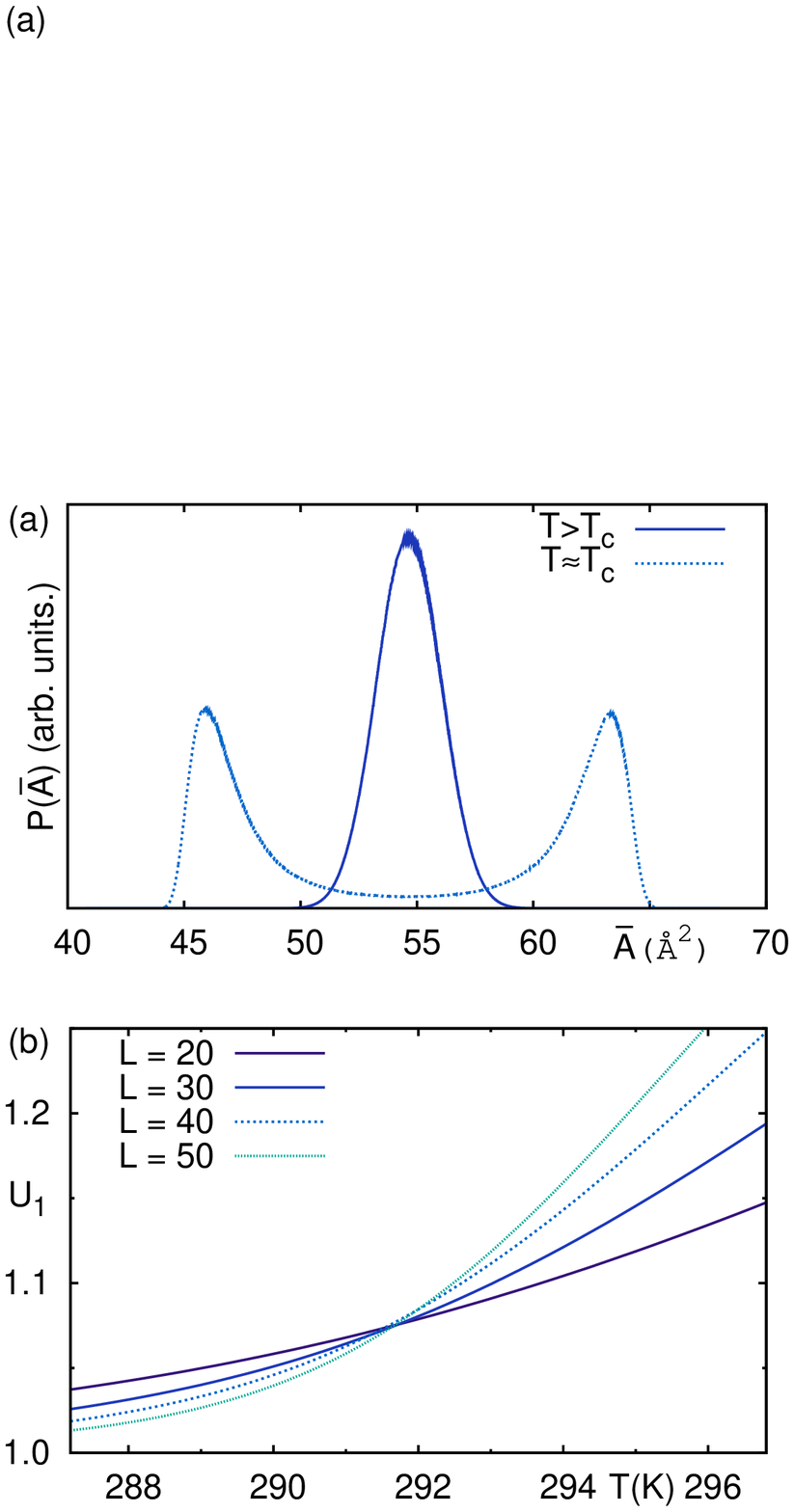} 
\caption{\label{probability-cumulant} Simulation results for DPPC obtained using 
the Pink model with \ahum{standard} parameters. (a) Probability distribution 
$P(\bar{A})$ of the cross-sectional area per molecule. Note that we have adopted 
the convention to plot the average area per lipid, $\bar{A} = 2A/N$, with $A$ 
given by \eq{eq:a}. At high temperature, irrespective of the value of $\Pi$, 
$P(\bar{A})$ is single-peaked corresponding to one phase (solid line). At low 
temperature, $P(\bar{A})$ becomes double-peaked provided $\Pi=\Pi_{\rm COEX}$, 
indicative of two-phase coexistence (dotted line). (b) Finite-size scaling 
analysis to locate the critical temperature $\Tc$. Plotted is the Binder 
cumulant $U_1$ as a function of temperature $T$ for different system sizes $L$. 
The intersection of the curves for different $L$ yields $\Tc$.}
\end{center}
\end{figure}

We first consider the main transition in a membrane consisting of DPPC 
lipids to settle a controversy when this system is being simulated using the 
Pink model. The acyl chains in DPPC consist of $M=16$ carbon atoms, and the 
experimentally obtained main transition temperature $T_{\rm DPPC} = 
314.0$~K~\cite{Ipsen1990}. However, simulations based on the Pink model could 
not detect a transition at this temperature~\cite{Laradji}. The latter 
simulations used the \ahum{standard} Pink parameters as listed in 
\tab{state_summary}, van der Waals coupling constant $J_0 = 0.710 
\times10^{-13}$~erg, and pressure $\Pi = 30$~dyn/cm. Hence the question arises 
as to why no transition could be detected. To answer this question we perform 
additional DPPC simulations using the Pink model, with the same parameters as in 
\olcite{Laradji}, but over a wider range in temperature and pressure. The 
picture that emerges is the following: At high temperature the distribution 
$\PA$ is always single-peaked (corresponding to one phase) for all value of the 
lateral pressure~$\Pi$. At low temperature, $\PA$ is doubled-peaked for a 
special value of the lateral pressure, $\Pi=\Pi_{\rm COEX}$, corresponding to 
two-phase coexistence [\fig{probability-cumulant}(a)]. Here, the left peak 
reflects the $gel$-phase, the right peak the $fluid$-phase. The numerical 
criterion to locate $\Pi_{\rm COEX}$ is to vary $\Pi$ until the fluctuation 
$\avg{A^2} - \avg{A}^2$ reaches a maximum~\cite{orkoulas.fisher.ea:2001}, with 
the thermal averages computed as $\avg{A^m} = \int A^m~\PA~dA$.

\begin{table}
\begin{ruledtabular}
\begin{tabular}{c c c c | c c c c c  c| c c c | c c c c}
     & & $M$ & & & $J_0$ & & $\Tc$ & & $\Pi_{\rm COEX}$ & & 
$T_{\rm m}$ & & & $J_{0}^{*}$ & & $\Pi_{\rm COEX}^{*}$ \\ \hline \hline
  DMPC & & 14 & & & 0.618 & & 270.3 & & 4.3  & & 296.9 & & & 0.690 & & 15.6 \\
  DPPC & & 16 & & & 0.710 & & 291.7 & & 4.6  & & 314.0 & & & 0.772 & & 18.1 \\
  DSPC & & 18 & & & 0.815 & & 321.5 & & 21.6 & & 327.9 & & & 0.833 & & 26.7 \\
\end{tabular}
\end{ruledtabular}
\caption{\label{critical-temp} Critical point parameters for three lipid 
species, with $M$ the number of carbon atoms in a single chain. We list the 
critical temperature $\Tc$ and coexisting pressure $\Pi_{\rm COEX}$ obtained in 
simulations of the Pink model using the \ahum{standard} value of the van der 
Waals coupling constant $J_0$. The resulting estimates of $\Tc$ are to be 
compared to the experimental melting temperatures $T_{\rm m}$: $\Tc$ clearly 
underestimates $T_{\rm m}$ in all cases. Instead, by using the Pink model with 
the re-tuned values $J_{0}^{*}$ proposed in this work, $\Tc$ coincides with 
$T_{\rm m}$, with corresponding critical pressure $\Pi_{\rm COEX}^{*}$ (coupling 
constants in units of $10^{-13}\, \rm{erg}$, temperatures in~K, and pressures 
in~dyn/cm).}
\end{table}

At the temperature $T=\Tc$ where the transition from a single to doubled-peaked 
distribution occurs, the system becomes critical. To locate the critical 
temperature a finite-size scaling analysis is performed, whereby we plot the 
Binder cumulant $U_1 = \avg{\Delta^2} / \avg{|\Delta|}^2$, $\Delta \equiv A - 
\avg{A}$, versus temperature~$T$ for different system sizes~$L$. In the 
thermodynamic limit
\begin{equation}\label{Binder}
 \lim_{L \to \infty} U_1 = \begin{cases}
  1     & T<\Tc, \\
  U_1^* & T=\Tc, \\
  \pi/2 & T>\Tc,
 \end{cases}
\end{equation}
while in systems of finite size, curves for different $L$ intersect at 
$T=\Tc$~\cite{binder:1981, binder:1981*c}. In \fig{probability-cumulant}(b), 
we show the result for DPPC obtained using the \ahum{standard} Pink model 
parameters: the data scale as expected, and from the intersection the critical 
temperature $\Tc$ can be accurately \ahum{read-off}.

The corresponding estimates of $\Tc$ as well as the coexistence pressures 
$\Pi_{\rm COEX}$ for three lipid species are collected in \tab{critical-temp}. 
For all lipid species considered, the computed critical temperature $\Tc$ is 
distinctly below the experimental melting temperature $T_{\rm m}$. In other 
words: if one simulates the Pink model at the experimental melting temperature 
$T_{\rm m}$, one is always inside the one-phase region, where $\PA$ is 
single-peaked! This, apparently, is the reason why no phase transition could be 
seen in previous studies~\cite{Laradji}. One possibility to get the proper value 
for the transition temperature, i.e.~such that $\Tc$ coincides with $T_{\rm m}$, 
is to re-tune the value of $J_0$. This has been done for the three lipid species 
by systematically changing the coupling constant $J_0$ using histogram 
reweighting and finite-size scaling. Our proposed values $J_{0}^{*}$ and 
corresponding pressures $\Pi_{\rm COEX}^*$ for the three lipid species are 
summarized in \tab{critical-temp}.

\begin{figure}
\begin{center}
\includegraphics[width=0.4\textwidth,angle=0]{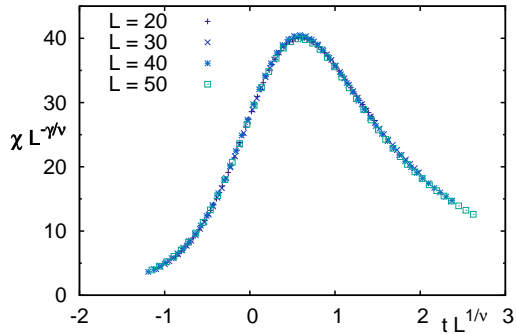} 
\caption{\label{scaled_susceptibility} Susceptibility scaling function $\chi \, 
L^{-\gamma/\nu}$ versus $t \, L^{1/\nu}$ for DPPC obtained using the 
\ahum{standard} Pink model. The data for different system sizes strikingly 
collapse using 2D Ising values for the critical exponents.}
\end{center}
\end{figure}

For completeness, we still confirm the universality class of the critical point, 
which for the Pink model is expected to be the one of the 2D Ising 
model~\cite{springerlink:10.1007/BF01295077}. To this end, we consider the 
susceptibility $\chi = \left( \avg{\Delta^2} - \avg{|\Delta|}^2 \right) / 
(k_{\rm B}TL^2)$~\cite{orkoulas.panagiotopoulos.ea:2000}, which diverges at the 
critical point $\chi \propto |t|^{-\gamma}$, $t=T/\Tc-1$, with critical 
exponent~$\gamma$. In systems of finite size, the divergence is rounded, but 
$\gamma$ can still be obtained using the standard finite-size scaling procedure 
of plotting $\chi \, L^{-\gamma/\nu}$ versus $t \, 
L^{1/\nu}$~\cite{newman.barkema:1999}, where $\nu$ is the correlation length 
critical exponent. Provided suitable values $\gamma,\nu,\Tc$ are used, data for 
different $L$ collapse. The result for DPPC is shown in 
\fig{scaled_susceptibility}, where the \ahum{standard} parameters of the Pink 
model were used. Indeed, by using the 2D Ising values $\{\gamma=7/4, \, 
\nu=1\}$, and $\Tc=291.7$~K of \tab{critical-temp}, an excellent data collapse 
is observed (similar good collapses are obtained for DMPC and DSPC also). The 
order parameter critical exponent has also been measured, and the 2D Ising value 
$\beta=1/8$ was confirmed (scaling plot not shown). Therefore, even though the 
Pink model is a 10-state model, its critical behavior remains in the 
universality class of the 2D Ising model. This further motivates the idea of 
reducing the $q=10$ states in the Pink model to an effectively two-state 
description as is frequently done~\cite{citeulike:6587773, citeulike:9172835, 
citeulike:9100609, springerlink:10.1007/BF01295077, citeulike:9234255}.

\subsection{modified Pink model with fewer states}

We now consider the effect of lowering the number of states in the Pink model. 
For this purpose, an appropriate number of intermediate states was removed, 
based on the maximum number of {\it gauche} bonds. In the \ahum{standard} 
10-state Pink model at most three {\it gauche} bonds are allowed. We now 
consider the case where at most two {\it gauche} bonds are permitted, by 
removing states $\alpha=8,9$ from \tab{state_summary}, yielding an 8-state model 
(to keep the total number of states constant the degeneracy of the removed 
states was added to the disordered state, but we emphasize that this correction 
is small). We apply our previous finite-size scaling analysis to the resulting 
8-state model for DPPC, using the \ahum{standard} value $J_0 = 0.710 \times 
10^{-13}$~erg. As expected, the critical point remains in the universality class 
of the 2D Ising model, but it is \ahum{shifted} to $\Tc=309.4$~K and $\Pi_{\rm 
COEX}=26.0$~dyn/cm. Similarly, by allowing at most one {\it gauche} bond, a 
5-state model is obtained. In this case, the DPPC critical point is located at 
$\Tc=351.5$~K and $\Pi_{\rm COEX}=87.7$~dyn/cm.

Therefore, lowering the number of states in the Pink model while leaving the 
other parameters untouched, one finds that both the critical temperature and 
pressure increase. This trend is consistent with the Pink model simulations of 
\olcite{Mouritsen2s} for DPPC performed at the experimental melting temperature 
$T_{\rm m}$ but with a lower number of states. In these simulations, 
hysteresis loops indicating a first-order transition are clearly visible around 
$T_{\rm m}$ for $q<6$. Indeed, as our scaling analysis shows, by lowering the 
number of states $q$, $\Tc$ eventually exceeds $T_{\rm m}$, resulting in a genuine 
phase transition at $T_{\rm m}$.

To conclude: lowering the number of states $q$ does not affect the universality 
class of the Pink model, which remains 2D Ising (provided $q \geq 2$, of 
course). Hence, the topology of the phase diagram remains the same, merely the 
critical point gets shifted. Depending on the parameters used, the main 
transition in the Pink model is either first-order ($T<\Tc$), or it is 2D Ising 
critical ($T=\Tc$). We do not claim that the main transition as observed in 
experiments necessarily conforms to this scenario (we return to this point in 
Section~V).

\subsection{Pink model with quenched disorder}

As a final illustration, we consider the main transition in a solid-supported 
membrane, which has received considerable attention in 
experiments~\cite{citeulike:10633341, citeulike:6591101, citeulike:9154736, 
citeulike:6090359, citeulike:6593910, citeulike:6590840}. A striking feature 
observed in one of these studies is the formation of coexisting gel and fluid 
domains that do not coalesce with time, but instead form a multi-domain 
structure that is stable over hours~\cite{citeulike:6090359}. To understand the 
stability of this structure is not trivial, due to the large amount of line 
interface it contains. Here we attempt to reproduce such a multi-domain 
structure within the framework of the Pink model. Our hypothesis is that the 
solid support onto which the membrane is deposited has a certain roughness. 
Since surface roughness is random and time independent, it constitutes a form of 
quenched disorder. We assume that this gives rises to regions on the surface 
where certain lipid tail conformations are preferred over others. We capture 
this effect in the Pink model by randomly labeling a fraction~$p$ of the lattice 
sites as \ahum{pinning sites}. At the pinning sites, the corresponding lipid 
chain is fixed into the ground state conformation. This extension is trivially 
incorporated into our MC simulations: we simply do not apply the MC move to 
pinning sites. We specialize to DSPC, using the \ahum{standard} Pink parameters 
of Tables~\ref{state_summary} and~\ref{critical-temp}.

\begin{figure}
\begin{center}
\includegraphics[width=0.4\textwidth,angle=0]{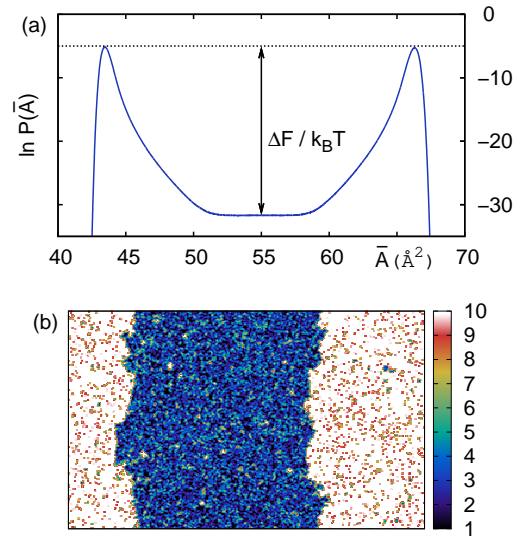}\

\caption{\label{FE-barrier} DSPC simulation results for $T=291$~K in the absence 
of quenched disorder. (a) The natural logarithm of $P(\bar{A})$ at $\Pi_{\rm 
COEX}$ and system size $L=50$. The barrier $\Delta F$ is related to the line 
tension via \eq{eq:lt}. (b) Typical snapshot of the bilayer with the lipids 
color-coded according to their conformational state for a $200 \times 300$ 
lattice. The snapshot was taken at cross-sectional area $\bar{A}=54.4$~\AA$^2$ 
chosen \ahum{between the peaks} of $P(\bar{A})$. A pronounced coexistence 
between a single gel and fluid domain is observed.}

\end{center}
\end{figure}

In \fig{FE-barrier}(a), we show $\ln \PA$ at $T = 291$~K and $\Pi_{\rm 
COEX}=-18.9$~dyn/cm in the absence of quenched disorder $(p=0)$. At this 
temperature, which is well below $\Tc$, the main transition is strongly 
first-order. Consequently, there is a significant line tension $\sigma$ between 
gel and fluid domains; the latter is related to the free energy 
barrier~\cite{binder:1982, billoire.neuhaus.ea:1994}
\begin{equation}\label{eq:lt}
 \Delta F \equiv k_{\rm B}T \ln \left( 
 P_{\rm max} / P_{\rm min} \right) = 2 \sigma L \quad,
\end{equation}
indicated by the vertical double-arrow in \fig{FE-barrier}(a). Here, $P_{\rm 
min}$ is the value of $\PA$ at the minimum \ahum{between the peaks}, while 
$P_{\rm max}$ denotes the average peak value. The physical motivation for 
\eq{eq:lt} is that, for cross-sectional areas \ahum{between the peaks}, the 
bilayer reveals a coexistence between two slab domains where the total interface 
length equals $2L$ [\fig{FE-barrier}(b)]. For DSPC, and assuming the lattice 
constant to be 1~nm, we obtain $\sigma \sim 1.1$~pN, which is compatible with 
experimental values~\cite{citeulike:10636141}.

\begin{figure}
\begin{center}
\includegraphics[width=0.4\textwidth,angle=0]{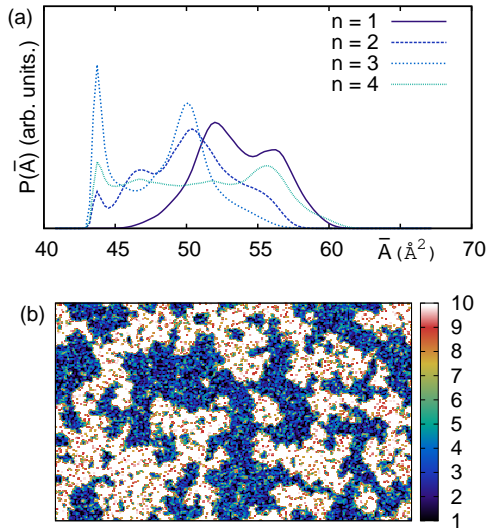}

\caption{\label{quenched-P} DSPC simulation results for $T=291$~K in the 
presence quenched disorder, with a fraction of pinning sites $p=0.03$. (a) 
Typical distributions $P(\bar{A})$ for 4 different samples of pinning sites, and 
system size $L=50$. In contrast to \fig{FE-barrier}(a), a first-order transition 
can no longer be identified. (b) Typical bilayer snapshot obtained at 
$\bar{A}=54.4$~\AA$^2$ for a $200 \times 300$ lattice. A stable structure of 
multi-domains is observed.}

\end{center}
\end{figure}

\begin{figure}
\begin{center}
\includegraphics[width=0.4\textwidth,angle=0]{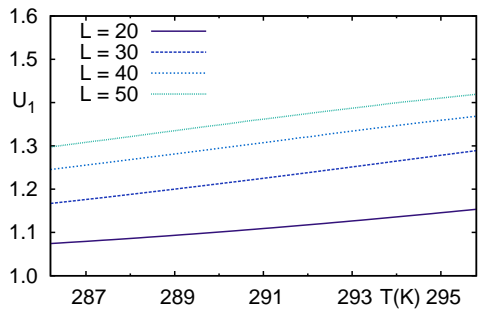}
\caption{\label{ave-Binder} The (disorder-averaged) Binder cumulant $U_1$ as a 
function of temperature $T$ for DSPC with a fraction $p=0.03$ of pinning sites. 
In contrast to \fig{probability-cumulant}(b), an intersection of the curves for 
different $L$ can no longer be identified. Instead, as the system size $L$ 
increases, $U_1 \to \pi/2$, indicative of the one phase region. For each system 
size, the disorder average comprised 200 samples of pinning sites.}
\end{center}
\end{figure}

Next, we consider a DSPC bilayer with a fraction $p=0.03$ of the lattice sites 
marked as \ahum{pinning sites}. In \fig{quenched-P}(a), we show distributions 
$\PA$ for $T=291$~K obtained for 4 different random positions (samples) of 
pinning sites. Even though the temperature is the same as in 
\fig{FE-barrier}(a), a unique double-peaked distribution can no longer be 
identified. In contrast, $\PA$ is strongly sample dependent, and a multitude of 
rather exotic shapes is revealed. This behavior is characteristic of systems 
that belong to the universality class of the 2D random-field Ising model 
(2D-RFIM)~\cite{citeulike:8864903}. Hence, by introducing the pinning sites, we 
have changed the universality class of the Pink model from ordinary 2D Ising 
toward 2D-RFIM (the pinning sites essentially correspond to a field of infinite 
strength acting at random locations).

There are two features of the 2D-RFIM universality class that are remarkably 
consistent with experimental results for the main transition in supported 
membranes. First of all, 2D-RFIM universality implies the absence of macroscopic 
coexistence between gel and fluid domains~\cite{imry.ma:1975, imbrie:1984, 
bricmont.kupiainen:1987, aizenman.wehr:1989}. Indeed, inspection of simulation 
snapshots [\fig{quenched-P}(b)] reveals an equilibrium multi-domain structure, 
that is highly anisotropic, strongly resembling experimental AFM 
images~\cite{citeulike:6090359}. A second (related) feature is that the 2D-RFIM 
has no true phase transition in the thermodynamic limit. In finite systems, 
there may be signs of a transition (or even several transitions; note that some 
of the distributions in \fig{quenched-P}(a) are triple-peaked), but they will be 
\ahum{smeared} over a wide temperature range, and do not persist in the 
thermodynamic limit. Precisely this behavior has also been reported in 
experiments~\cite{citeulike:6090359, citeulike:6591101}. Simulation evidence 
that the pinning sites prevent a sharp transition in the thermodynamic limit 
follows from the (disorder-averaged) Binder cumulant $U_1={\left[\left \langle 
\Delta^{2} \right \rangle \right]}/ {[{\left \langle \left |\Delta \right | 
\right \rangle}^2]}$, where $[\cdot]$ denotes an average over different samples 
of pinning sites. As shown in \fig{ave-Binder}, $U_1 \to \pi/2$ as $L$ 
increases, consistent with only a single phase.

\section{Conclusion}

In this paper, the main phase transition in single-component phospholipid 
membranes was investigated using the Pink model. Our simulations of the pure 
membrane (i.e.~without quenched disorder) confirm the formation of macroscopic 
gel and fluid domains below a critical temperature $\Tc$, and at the coexistence 
pressure $\Pi_{\rm COEX}$. We also demonstrated that, using the accepted values 
of the Pink model parameters, $\Tc$ falls below experimentally measured 
transition temperatures. This explains why no phase transition was detected at 
the experimental transition temperature in the simulations of \olcite{Laradji}. 
To resolve this issue, we propose that the strength of the van der Waals 
coupling in the Pink model be increased. By using the values proposed in this 
work, $\Tc$ of the Pink model {\it in the thermodynamic limit} coincides with 
the experimental main transition temperature. In addition, finite-size scaling 
was applied to confirm the universality class of the critical point in the Pink 
model, which was shown to be 2D Ising. This result holds irrespective of the 
number of conformational states $q$ (as long as $q \geq 2$, of course). Hence, 
to capture the generic features of membrane phase behavior, a highly-detailed 
model is not always needed (which is consistent with the findings of 
\olcite{citeulike:10648601}).

We have also used the Pink model to describe the main transition in the presence 
of quenched disorder, which may arise in case the membrane is deposited on a 
rough support. Assuming that this induces regions in the membrane where certain 
tail conformations become preferred, the universality class changes toward that 
of the 2D {\it random-field} Ising model. In the presence of quenched disorder, 
the Pink model reveals a stable multi-domain structure, and the absence of a 
sharp transition; these findings are indeed consistent with some of the 
experimental observations.

Although the Pink model (both the pure version and the one containing quenched 
disorder) seems well suited to describe the main transition, we wish to end with 
a warning. By using the Pink model, one inevitably casts the main transition 
into the Ising universality class. This may not be entirely appropriate, as the 
main transition is essentially a melting transition leading to the formation of 
nematic chain order. A liquid-crystal model may therefore be more suitable, such 
as the Maier-Saupe approach followed in \olcite{citeulike:9100874}. If, indeed, 
the main transition occurs close to a critical point \cite{citeulike:6587773, 
citeulike:10561379} it may well be necessary to replace the discrete set of 
states of the Pink model by a continuous one \cite{citeulike:4074809}. In that 
case, one enters the regime of Heisenberg-type models (which, provided certain 
conditions are met, do support 2D phase transitions \cite{physrevlett.89.285702, 
physrevlett.88.047203, citeulike:5202797}). The investigation of the main 
transition in terms of such a continuous model could be an interesting topic for 
future work.

\acknowledgments

This work was supported by the {\it Deutsche Forschungsgemeinschaft} (Emmy 
Noether~VI~483 and the SFB-937).

%% USE BIBTEX !!!
\bibliography{refs,refs_VINK}

\end{document}